\documentclass[aps,prb,twocolumn,showpacs,superscriptaddress]{revtex4-1}\begin{tiny}\end{tiny}
\usepackage{longtable}
\usepackage{amsfonts}
\usepackage[dvipdfm]{graphicx}
\usepackage{setspace}

\begin{document}

\title{Heat Conduction by Phonons across a Film}

\author{Philip B. Allen}
 \email{philip.allen@stonybrook.edu}
 \affiliation{Physics and Astronomy Department, Stony Brook University, Stony Brook, NY 11794-3800, USA}

\date{\today}


\begin{abstract}
Quasiparticle theory gives a local relation between heat current and temperature gradient,
provided the quasiparticle mean free path is smaller than the scale of variation of temperature.
When mean free paths are comparable to sample size, the relation becomes non-local.
This non-local relation is formulated for phonon carriers; an explicit form is found in the approximation
where current relaxation rates are replaced by quasiparticle relaxation rates.  A quasiparticle
definition of local temperature is offered.  To extract the spatial
variation of the temperature gradient requires inverting the non-local relation.
A variational principle is constructed.  The heat current is bounded above when evaluated for
a trial temperature gradient.  The true temperature gradient is the one which minimizes
heat current.  The simplest variational approximation (a constant temperature gradient)
gives an approximate formula for the
thickness dependence of the heat transport across a film whose thickness is comparable to
mean free paths of phonons.  The formula interpolates between ballistic and diffusive limits. 
\end{abstract}

\maketitle


\section{Introduction}

This paper examines heat transport in small systems where boundary and size effects play an important role.  Only steady state situations
are considered, where the heat current $j$ is independent of time.  The systems considered are spatially homogeneous films,
with inhomogeneous temperature profiles perpendicular to the film, in the direction of current flow.

Heat transport in such systems depends on the quality of interfaces, and is often studied with an aim of diminishing
the transport by interface scattering \cite{Dames}.  The present paper discusses an idealized model in the opposite
limit, where interfaces transmit phonons smoothly from a thermalized bath at one temperature $T(x=0)=T_0$ to another
thermalized bath at $T(x=L)=T_L$ with $T_0>T_L$.

\section{Phonon Quasiparticle Theory}

The premises of quasiparticle theory are the following.  (1) Propagating vibrational normal modes exist.  
``Existence'' signifies that they live long enough and propagate far enough that their wavevector $\vec{Q}$ 
and frequency $\omega_Q$ can be measured to decent accuracy.  The short-hand $Q$ is used to denote both 
the wavevector $\vec{Q}$ and the branch index $s$.  (2) The dynamics is close enough to harmonic that to first approximation, 
the thermal properties are describable by $N_Q(\vec{r},t)$, the mean occupation of mode $Q$.  (3) The dynamics of $N_Q$ is 
governed by a closed evolution equation, and this equation can describe the evolution towards a thermal equilibrium
 Bose-Einstein distribution $n_Q = 1/(\exp(\hbar\omega_Q/k_B T)-1)$.
 
 As an example of premise (2), consider the heat current density in the $\hat{x}$ direction, $j_x(\vec{r},t)$.  It is given by
\begin{equation}
j_x(\vec{r},t) = \frac{1}{\Omega}\sum_Q \hbar\omega_Q v_{Qx} N_Q(\vec{r},t),
\label{eq:j}
\end{equation}
where $\Omega$ is the volume of the sample, and $v_{Qx}=\partial \omega_Q /\partial Q_x$ is the group velocity.
By time-reversal invariance, the heat current vanishes in thermal equilibrium, when $N_Q=n_Q$,
including the case of local equilibrium where $n_Q$ varies in space through a spatially
varying temperature $T(\vec{r})$.   Let the sample be reasonably large,
and the driving by external heat sources reasonably weak.  It is then plausible to assume that a local temperature
$T(\vec{r})$ can be defined.  This assumption will be discussed further later on.  If the weak driving is constant in time,
the system reaches a steady state, characterized by a slowly-varying local temperature.  If, as is commonly observed, the Fourier law 
$j_x = -\kappa_{xx} \nabla_x T$ correctly describes the heat flow, then it should follow from Eq.(\ref{eq:j}) 
under a first-order expansion in the temperature gradient.  The conductivity $\kappa_{xx}$ could
depend on the local temperature, but should not depend on the details of sample size and shape.
However, at low $T$, where long wavelength modes may have mean free paths comparable to the system size,
modifications should be expected, and the relation between current and temperature gradient will become
non-local.

\section{Model system: the Gentle Slab}

Consider a homogeneous film of thickness $L$, where one side (at $x=0$) is held at temperature $T_0$ and the other side (at $x=L$) is held at
temperature $T_L < T_0$.  A steady heat flow will appear down the temperature gradient.  In an actual experiment of this type, there are
difficulties with quasiparticle theory.  First, the true normal modes obey boundary conditions at $x=0$ and $x=L$ 
which are more complicated than
the periodic boundary conditions used to describe quasiparticles with well-defined wavevectors and group velocities.  
The actual normal modes are standing waves; the traveling waves (wave-packets) 
that transport heat are non-stationary combinations of normal modes.  
Second, heat must be injected at $x=0$
and removed at $x=L$; this process may be complicated to model.  Therefore, a simpler model is needed, which I call a ``gentle slab.''

The ``gentle slab'' is a homogeneous film, with an interior portion $0 < x < L$ where the conductivity is studied.  The interior
part will be referred to as the ``system.''  There are two exterior heat baths.
The exterior part with $x < 0$,
held at temperature $T_0$, serves as the high temperature heat bath.  The exterior part with $x > L$, held at temperature $T_L$,
serves as the low temperature heat bath. This situation is easily simulated numerically, using ``thermostats'' \cite{thermostat} in the
$x<0$ and $x>L$ regions.  This configuration provides a model system for which the non-local relations can be carefully studied.

\section{Ballistic Transport} 

Debye \cite{Debye}  was the first to understand that a perfect harmonic crystal should carry heat ballistically.  The ``gentle slab'' gives
a nice model.  An effective conductivity can be defined by $j_x = - \kappa_{\rm eff}(T_L-T_0)/L$, but the effective conductivity
$\kappa_{\rm eff}$ depends on the system size, violating the normal understanding of the Fourier law.    In the interior ($0<x<L$), phonons with $v_{Qx}>0$ 
have propagated in from the region $x<0$ and thus have $N_Q = n_Q(T_0)$, and those with $v_{Qx}<0$ have propagated in 
from the right, and have $N_Q = n_Q(T_L)$.  The heat current is thus
\begin{equation}
j_x = \frac{1}{\Omega}\sum_Q^{v_{Qx}>0} \hbar\omega_Q v_{Qx}[n_Q(T_0)-n_Q(T_L)],
\label{eq:bal}
\end{equation}
where the symmetry $\omega_{-Q}=\omega_Q$ and $v_{-Qx}=-v_{Qx}$ has been used to simplify the sum. 
This can be evaluated in the Debye model, where there are three acoustic branches per atom, all having
$\omega_Q=v|\vec{Q}|$, and all having the same velocity $v$.  The answer is
\begin{equation}
j_x = (3v/4)[u(T_0)-u(T_L)]
\label{eq:flow}
\end{equation}
where $u(T)$ is the thermal vibrational energy density.  At low $T$, this
has the value $u(T)=(3\hbar\omega_D/5\Omega)(\pi k_B T/\hbar\omega_D)^4$.  We get the result
$j_x \propto T_0^4 - T_L^4$ familiar from elementary texts for radiative heat transport.  If we let $T$ denote
the mean temperature $(T_0+T_L)/2$ and if the temperature difference $\Delta T = T_0 - T_L$ is small compared to $T$,
then Taylor expansion is accurate, and we obtain
\begin{equation} 
j_x = \frac{1}{\Omega}\sum_Q^{v_{Qx}>0}  \hbar\omega_Q v_{Qx}L \frac{\partial n_Q}{\partial T} \left[ \frac{T_0 - T_L}{L} \right].
\label{eq:bal}
\end{equation}
The last factor, in square brackets, is the negative temperature gradient, so we get an effective conductivity that depends on $L$.
In a Debye approximation, this becomes
\begin{equation}
\kappa_{\rm eff}=\frac{1}{4}C\overline{v}L,
\label{eq:}
\end{equation}
reminiscent of the usual textbook formula $\kappa = (1/3)C\overline{v}\ell$, where $C$ is the specific heat and $\ell$ is the average
mean free path.  In the ballistic case, the mean free path is replaced by the film thickness, and a revised angle average changes
$1/3$ to $1/4$.

\section{Peierls-Boltzmann Theory and the Definition of Temperature}

In the ballistic case just described, temperature is ill-defined in the interior.  Even when the mean free path
of the heat carrier becomes shorter than the system size, it is not obvious how temperature should be defined,
as Cahill {\it et al.} \cite{Cahill} have emphasized  Simulations are generally classical,
since it is difficult to model with the true quantized vibrations.  
The classical vibrational problem has a natural definition of local temperature, namely {\it via} the
mean kinetic energy ($<KE>=3k_B T(x)/2)$ of atoms with coordinate $x$.   For the ballistic gentle slab,
the result is a discontinuous temperature, $T_0$ when $x<0$, $T_L$ when $x>L$, and the mean
temperature $T$ everywhere in between. 

The evolution equation for the phonon occupation $N_Q$ is the Peierls Boltzmann equation. \cite{Peierls}  
Although not exact, it is the correct quasiparticle
theory, likely to remain valid up to the point where scattering is sufficiently strong to destroy quasiparticles. \cite{Sun}  This equation has been
confirmed by derivation in the linear response limit, from Kubo formulas. \cite{Krumhansl}  In a 3-d simulation using a fairly small cell size, 
the gradients may be large enough to involve non-linear effects. \cite{Mahan}  The Peierls-Boltzmann equation contains in principle the leading
non-linear effects also.  In steady state ($\partial N_Q/\partial t = 0$) the form of the equation is
\begin{equation}
\left( \frac{\partial N}{\partial t} \right)_{\rm drift} =   - v_{Qx}\frac{\partial N_Q}{\partial x} = 
- \left(\frac{\partial N_Q}{\partial t}\right)_{\rm collision},
\label{eq:fullPB}
\end{equation}
where the collision term on the right is a complicated non-linear sum (or integral) over other phonon states $N_{Q^{\prime}}$.
It is tedious to solve the full integral equation, even when appropriately linearized.  Therefore, it is
common to approximate the collision term by a simplified linear and local-in-$Q$ approximation,
\begin{equation}
(\partial N_Q/\partial t)_{\rm collision} \rightarrow -(N_Q - n_Q)/\tau_Q,
\label{eq:relaxPB}
\end{equation}
where $\tau_Q$ is a phonon relaxation time.  Linearizing also the left side of Eq.(\ref{eq:fullPB}),
we could write 
\begin{equation}
(N_Q - n_Q)=v_{Qx}\tau_Q^{\rm tr}\partial n_Q/\partial x.
\label{eq:tauQ}
\end{equation}
The ``exact'' phonon relaxation time $\tau_Q^{\rm tr}$ is defined by Eq.(\ref{eq:tauQ}) if $N_Q-n_Q$ is the
exact solution of the linearized version of Eq.(\ref{eq:fullPB}).  It is common, and fairly accurate except at low $T$,
to take as the value of $\tau_Q$, the one predicted by Eq.(\ref{eq:fullPB}) when
all phonons $Q^\prime$ are forced to be in equilibrium ($N_{Q^\prime}=n_{Q^\prime}$) except when $Q^\prime$ equals $Q$.  
This is the ``quasi-particle relaxation time,'' and 
agrees with $-2\rm{Im}\Sigma(Q,\omega)$, the imaginary part of the phonon self-energy, in lowest
order perturbation theory.  The quasi-particle
relaxation time approximation tends to underestimate the thermal conductivity.  
The reason is that conductivity does not depend only on
quasiparticle relaxation; what is really needed is current relaxation contained in $1/\tau_Q^{\rm tr}$.
When $N_Q \ne n_Q$, all collisions involving mode $Q$
tend to relax the quasiparticle population of state $Q$.
However, some collisions do not relax the current, especially the ``N-processes'' (N=Normal, U=Umklapp.)  The full 
Eq.(\ref{eq:fullPB}) correctly relaxes the current, a slower process than thermalizing the quasiparticle distribution. 

The full Peierls-Boltzmann equation has a nice property, not often mentioned, namely, satisfaction of the Boltzmann ``H-theorem.''
\cite{Pauli}   Entropy, unlike energy, is not rigorously defined except in equilibrium.  Nevertheless, there are two arguments that show
that the quantity 
\begin{equation}
S = k_B\sum_Q[(N_Q +1)\ln(N_Q +1)-N_Q \ln N_Q].
\label{eq:S}
\end{equation}
 is an appropriate entropy for quasiparticles (if they exist), both in thermal equlibrium (when the distribution function $N_Q$ 
 is replaced by the equilibrium value $n_Q$), and when driven out of equilibrium (assuming that quasiparticles
 continue to exist.)   One argument is the fact that this equation follows by counting available states
without assuming equilibrium. \cite{Landau}  The other argument is that one can use the Peierls-Boltzmann equation in
its full form (Eq. \ref{eq:fullPB}) to compute the rate of change $dS/dt$ of Eq.(\ref{eq:S}) caused by collisions.  This permits the
conclusions (1) that this version of entropy never decreases in time, and (2) that the only way that this version of $S$ can be stationary
under collisions is by having $N_Q$ equal a local equilibrium Bose-Einstein function $n_Q$.  This corresponds to the usual equilibrium
harmonic oscillator entropy, except it is allowed to have spatial variation of $T$.
Let us regard Eq.(\ref{eq:S}) as defining a local $\vec{r}$
and $t$-dependent entropy.  Next, define the corresponding $\vec{r}$ and $t$-dependent energy
\begin{equation}
U = \sum_Q \hbar\omega_Q (N_Q + 1/2).
\label{eq:U}
\end{equation}
These quasiparticle equations permit a ``local steady-state quasiparticle temperature'' to be defined
{\it via} the steady-state non-equilibrium quasiparticle distribution function,
\begin{equation}
\frac{1}{T(\vec{r})} \equiv \frac{k_B}{3N}\sum_Q \frac{\partial S}{\partial N_Q} \left[\frac{\partial U}{\partial N_Q}\right]^{-1} 
\label{eq:}
\end{equation}
Solving Eq.(\ref{eq:fullPB}) for $N_Q$ might then allow $T(\vec{r})$ to be properly defined.  In the next section, something
like that will be done, except in a circular way.  It will be assumed that the solution $N_Q$ contains the function $T(\vec{r})$,
and this will allow $N_Q$ to be computed and $T(\vec{r})$ to be defined.

\section{Non-local conductivity}

Both Chen's book \cite{Chen} and Zhang's book \cite{Zhang} have nice discussions of size effects under parallel transport in a thin film, obtained by solving Eq.(\ref{eq:fullPB},\ref{eq:relaxPB}).  
The application to perpendicular transport is presented here.  Write the steady-state solution for
the distribution function as $N_Q(x) = n_Q(x) + \Phi_Q(x)$, where the equilibrium part $n_Q$ depends on $x$ through
some definition of local temperature $T(x)$.  Then Eq.(\ref{eq:relaxPB}) becomes
\begin{equation}
\frac{\partial \Phi_Q}{\partial x} + \frac{\Phi_Q}{\ell_Q} = -\frac{\partial n_Q}{\partial x}
\label{eq:PBE}
\end{equation}
where $\ell_Q = v_{Qx}\tau_Q$.  The boundary conditions are  $\Phi_Q(x=0) = 0$ if $\ell_Q > 0$, and 
$\Phi_Q(x=L) = 0$ if $\ell_Q < 0$.  The solution is
\begin{eqnarray}
\Phi_Q &=& -\int_0^x dx^\prime e^{-(x-x^\prime)/\ell_Q} \frac{\partial n_Q}{\partial x^\prime} \ \  \rm {if} \ \ell_Q >0  \nonumber \\
&=&  -\int_L^x dx^\prime e^{-(x-x^\prime)/\ell_Q} \frac{\partial n_Q}{\partial x^\prime} \ \  \rm {if} \ \ell_Q <0.
\label{eq:PBEsol}
\end{eqnarray}
Using Eq.(\ref{eq:j}), $j=\sum \hbar\omega_Q v_{Qx}\Phi_Q$, the current can be written as
\begin{equation}
j_x(x) = -\int_0^L dx^\prime  \kappa(x,x^\prime)  \frac{\partial T(x^\prime)}{\partial x^\prime},
\label{eq:Fournonloc}
\end{equation}
where the non-local conductivity is
\begin{equation}
\kappa(x,x^\prime) = 
\frac{1}{\Omega}\sum_Q^{v_{Qx}>0} \hbar\omega_Q v_{Qx} \frac{\partial n_Q(x^\prime)}{\partial T} e^{-|x-x^\prime|/\ell_Q}
\label{eq:kappanonloc}
\end{equation}
%

%
%
%
%

Two limiting cases can be checked.  First, if relevant mean free paths $|\ell_Q|$ are small on the scale of $L$ and on the scale of
$T/|dT/dx|$, then one is in the ``diffusive limit'' where the usual Fourier law applies.  The factor $e^{-|x-x^\prime|/\ell_Q}$
can be replaced by $2\ell_Q \delta(x-x^\prime)$, so 
\begin{equation}
\kappa(x,x^\prime)\rightarrow \frac{2}{\Omega}\sum_Q^{v_{Qx}>0} \hbar\omega_Q v_{Qx} \frac{\partial n_Q }{\partial T}
\delta(x-x^\prime) \approx \frac{1}{3}C\bar{v}\ell \delta(x-x^\prime)
\label{eq:kappa}
\end{equation}
which gives the usual Fourier law.  Second, if the relevant mean
free paths are all long compared with $L$, then $\kappa(x,x^\prime)$ has for spatial dependence, only
the factor $\partial n_Q (x^\prime)/\partial T$.  Multiplying by $\partial T/\partial x^\prime$ and integrating $dx^\prime$ over
the sample thickness gives exactly the ballistic result, Eq.(\ref{eq:bal}).

\section{Entropy}

The question not yet addressed, is to find the temperature profile $T(x)$ when a current $j_x$ is flowing.  Since there is no
driving except external to the interval $0<x<L$, 
the steady state current must be constant throughout the gentle slab; that is, $j_x(x)$ is independent
of $x$.  In the first (diffusive)  limit, this tells us that the gradient $dT/dx$ is necessarily constant, going smoothly from $T_0$ at $x=0$ to 
$T_L$ at $x=L$.  In the second (ballistic) limit, one does not need to specify a temperature in the interior.  Only the end temperatures are
required.  
The phonons are
unaware of temperature, once launched in the heat baths.  Temperature is both ill-defined and irrelevant.
The situation becomes interesting in the crossover regime, where
the intermediate temperature profile needs specification.  Equation (\ref{eq:Fournonloc}) with the left side constant becomes an
integral equation that must be solved for $dT(x)/dx$.  The operator $\kappa(x,x^\prime)$ is real symmetric in the classical limit where
$dn_Q/dT = k_B /\hbar\omega_Q$ is independent of $x$.  In the quantum case, for small temperature differences $|T_0 - T_L | \ll T$, the
$x^\prime$-dependence of $\partial n_Q /\partial T$ becomes ignorable, which also makes
 $\kappa(x,x^\prime)$ real symmetric.  The second law of thermodynamics (non-decreasing entropy) requires
 $\kappa(x,x^\prime)$ to be a non-negative operator, allowing variational approximation to the inversion.
 
If, in time $\Delta t$, an amount of heat $\Delta Q$ enters the ``system'' at the $x=0$ boundary and leaves through the
$x=L$ boundary, the total entropy increase is $\Delta S = \Delta Q (1/T_L-1/T_0)$.  The left reservoir loses entropy
$Q/T_0$ and the right reservoir gains $Q/T_L$.  The ``system'' itself is in steady state, and has no entropy change,
\begin{eqnarray}
&&0=\frac{dS}{dt} = \left(\frac{dS}{dt}\right)_{\rm drift} + \left(\frac{dS}{dt}\right)_{\rm collision} \\
&&=k_B \sum_Q \ln\left( \frac{N_Q+1}{N_Q} \right) \left[ \left( \frac{dN_Q}{dt} \right)_{\rm drift}+
\left( \frac{dN_Q}{dt} \right)_{\rm collision} \right] \nonumber
\label{eq:dSdt}
\end{eqnarray}
The separate parts, $(dS/dt)_{\rm drift}$ and $(dS/dt)_{\rm collision}$ cancel.  The ``drift'' part is negative, 
exactly cancelling the net entropy increase of the baths, as will be shown below.  
Therefore, the source of the entropy increase of the
whole (system plus baths) can be identified with the collisions in the system.  A 
microscopic version of the total entropy increase is 
\begin{eqnarray}
&&\left(\frac{dS}{dt}\right)_{\rm tot} = -\frac{k_B A}{\Omega} \int_0^L dx
 \sum_Q \ln\left( \frac{N_Q+1}{N_Q} \right) \left( \frac{dN_Q}{dt} \right)_{\rm drift} \nonumber \\
&&=\frac{k_B A}{\Omega}\int_0^L dx \sum_Q \ln\left(\frac{1+n_Q +\Phi_Q}{n_Q +\Phi_Q}\right) \frac{\partial n_Q}{\partial T}v_{Qx}
\frac{\partial T}{\partial x},
\label{eq:dStotdt}
\end{eqnarray}
where $A$ is the area of the film.  Again assuming small temperature difference so that the steady state is not far from equilibrium
($|\Phi_Q| \ll n_Q$), the logarithm can be Taylor expanded:
\begin{equation}
\ln\left(\frac{1+n_Q +\Phi_Q}{n_Q +\Phi_Q}\right) \approx \frac{\hbar\omega_Q}{k_B T}-\frac{\Phi_Q}{n_Q (n_Q +1)}
\label{eq:log}
\end{equation}
The first term on the right of Eq.(\ref{eq:log}) gives vanishing entropy increase from Eq.(\ref{eq:dStotdt}).  The second term simplifies
because $dn_Q/dT = n_Q (n_Q+1)(\hbar\omega_Q/k_B T^2)$.  The result is
\begin{eqnarray}
&&\left(\frac{dS}{dt}\right)_{\rm tot} =  -\frac{A}{\Omega T^2} \int_0^L dx \sum_Q \Phi_Q \hbar\omega_Q v_{Qx}\frac{dT(x)}{dx}  \nonumber \\
&&=-\frac{A}{T^2}\int_0^L dx  j_x(x)\frac{dT(x)}{dx} =j_x A\left(\frac{1}{T_L}-\frac{1}{T_0}\right)   \nonumber \\
&&= \frac{A}{T^2}\int_0^L dx  dx^\prime \frac{dT(x)}{dx} \kappa(x,x^\prime) \frac{dT(x^\prime)}{dx^\prime}\ge0 
\label{eq:pos}
\end{eqnarray}
The second identity of the second line of Eq.(\ref{eq:pos}) follows from the first because the current $j_x$ is independent of $x$;
$j_x A$ is, of course, the rate of heat transfer $dQ/dt$.
This line confirms that the negative of the rate of change of entropy of the ``system'' due to drift is indeed the same as the rate of increase of the
total entropy (``system'' plus boundaries).  The third line shows that $\kappa(x,x^\prime)$ has all diagonal matrix elements
positive: $(g,\kappa g)$ is positive for any choice $g=dT/dx$ of temperature gradient.  
The only way $dS_{\rm tot}/dt$ can be zero is for the system to be in equilibrium; that is, the
temperature gradient must be zero.  We can now exploit the fact that $\kappa(x,x^\prime)$ is a positive operator.

\section{A Variational Principle}

Our aim is to compute the temperature gradient when there is a steady heat flow, and ultimately the dependence of heat current
on temperature difference.  Eq.(\ref{eq:pos}) can be written as
\begin{equation}
(g,\kappa g) = j_x (T_0 - T_L) \ge \frac{(h,\kappa g)^2}{(h,\kappa h)}=\frac{(h,j_x)^2}{(h,\kappa h)}
\label{eq:Schwartz}
\end{equation}
where $g$ is the true temperature gradient, responsible for the actual heat current $j_x$, and $h$ is any trial 
temperature gradient.    The inequality becomes an equality when $h$ is proportional to $g$.
Eq.(\ref{eq:Schwartz}) uses the Schwartz inequality, $\langle a|a\rangle  \langle b|b\rangle\ge\langle a|b\rangle^2$, valid for any inner product.
Specifically, the positivity of $\kappa$ allows an inner product $\langle a|b\rangle$ to be defined as $(a,\kappa b)$.
The true solution $h = cg$ is found by varying $h$ until it maximizes the right hand side of Eq.(\ref{eq:Schwartz}).
It is a variational principle for the heat current across a film.  It is reminiscent of the Kohler variational principle
which is useful for finding good approximations to the linearized Boltzmann transport equations in homogeneous
media \cite{Ziman}.  Like the Kohler principle, its interpretation is that the true solution is the one that maximizes
entropy production; any approximate solution underestimates entropy production and conductivity.  The new
principle applies to the exact Peierls-Bolzmann equation for the gentle slab, as well as to the approximate version
obtained from relaxation-time approximation.  It should surely also apply to the Bloch-Boltzmann (electron) equation for the
same model, and probably for the electrical conductivity problem as well as heat conductivity.  Probably it can
be extended to more complicated models.

Using the fact that $j_x$ is independent of $x$, the numerator on the right of Eq.(\ref{eq:Schwartz})
is just $j_x (h,1)$, the variational principle can be re-written as
\begin{equation}
j_x \le (T_0 - T_L) \frac{(h,\kappa h)}{(h,1)^2}
\label{eq:var}
\end{equation}
For the explicit approximate version of Eq.(\ref{eq:kappanonloc}), the variational answer can be written
\begin{equation}
j_x \le j_x^{(p)},  \nonumber
\end{equation}
\begin{equation}
j_x^{(p)} = \frac{(T_0 - T_L)}{L}\frac{2}{\Omega} \sum_Q^{v_{Qx}>0} \hbar\omega_Q v_{Qx}\ell_Q \frac{dn_Q}{dT}F^{(p)}(\ell_Q/L),
\label{eq:var2}
\end{equation}
\begin{equation}
F^{(p)}(\ell_Q/L)=\frac{L}{2\ell_Q}
 \frac{\int_0^{L} dx\int_0^L dx^\prime e^{-|x-x^\prime|/\ell_Q}h^{(p)}(x)h^{(p)}(x^\prime) } {\left[\int_0^{L} dx h^{(p)}(x) \right]^2},
 \label{eq:var3}
\end{equation}
where $(p)$ indicates a $p$-th generation trial gradient $h^{(p)}$ with variational parameters $\{\alpha_1, .. \alpha_p\}$.
The symmetry of the ``gentle slab'' model causes $\kappa(x,x^\prime)$ to be invariant under $(x,x^\prime)
\rightarrow (-x,-x^\prime)$.  The solution will have therefore ``left-right'' symmetry $g(L-x)=g(x)$.
Choosing the trial function $h(x)$ also to have this symmetry, the integrations in Eq.(\ref{eq:var3}) simplify.
%
%
The simplest trial gradient is, of course, a constant, $h^{(0)}(x)=1$ (the actual value,
$(T_L-T_0)/L$, is irrelevant because it cancels with the denominator.)  This gives an
expression for $F^{(0)}$,
\begin{equation}
F^{(0)}=1-\left( \frac{1-e^{-L/\ell_Q}}{L/\ell_Q}\right)
\label{eq:F}
\end{equation}
%
%
\begin{eqnarray}
j_x \le j_x^{(0)} &=& \frac{(T_0 - T_L)}{L}\frac{2}{\Omega}\sum_Q^{v_{Qx}>0}\hbar\omega_Q v_{Qx}\ell_Q \frac{dn_Q}{dT} \nonumber \\
&\times& \left[ 1-\left( \frac{1-e^{-L/\ell_Q}}{L/\ell_Q}\right)\right]
\label{eq:lova}
\end{eqnarray}
For short mean free paths $\ell_Q \ll L$, the correction factor $F^{(0)}$ is 1 and the current is the usual
bulk conductivity.  For long mean free paths $\ell_Q \gg L$, the factor is $F^{(0)}=L/2\ell_Q$, and the 
current is the ballistic result, Eq.(\ref{eq:bal}).  So the expression is correct in both limits, and gives a
smoothly interpolating upper bound in between.

It seems unlikely that an improved variational correction can be written explicitly.  In actual
computational theory, if the non-local $\kappa(x,x^\prime)$ is constructed, numerical inversion
would be preferable to variational approximation.  But the interpolation formula given
here should be reasonably accurate and has the virtue of simplicity.

\section{acknowledgements}
This work was supported in part by DOE grant No. DE-FG02-08ER46550.


\begin{thebibliography}{99}
\bibitem{Dames} C. Dames and G. Chen, Thermal Conductivity of Nanostructured
Thermoelectric Materials, CRC Handbook, edited by M. Rowe
(Taylor and Francis, Boca Raton, FL, 2006), Ch. 42.
\bibitem{thermostat} P. H\"unenberger, Adv. Polym. Sci. {\bf 173}, 105 (2005).
\bibitem{Debye} P. Debye, {\it Equation of state and quantum hypothesis, with an appendix
on thermal conductivity} (German text), in {\it Lectures on the Kinetic
Theory of Matter and Electricity}, M. Planck, {\it et al.}, editors,
Leipzig, Teubner, 1914, p. 19-60.
\bibitem{Cahill} D. G. Cahill, W. K. Ford, K. E. Goodson, G. D. Mahan, A. Majumdar, H. J. Maris, R. Merlin, and S. R. Phillpot,
J. Appl. Phys. {\bf 93}, 793 (2003).
\bibitem{Peierls} R. Peierls, Ann. Phys. {\bf 3}, 1055 (1929).
\bibitem{Sun} T. Sun and P. B. Allen, Phys. Rev. B {\bf 82}, 224305 (2010).
\bibitem{Krumhansl} C. Horie and J. A. Krumhansl, Phys. Rev. {\bf 136}, A1397 (1964).
\bibitem{Mahan} G. D. Mahan and F. Claro, Phys. Rev. B {\bf 38}, 1963 (1988).
\bibitem{Pauli} W. Pauli, 1928 {\it Sommerfeld Festschrift}, Leipzig, 1928.
\bibitem{Landau} L. D. Landau and E. M. Lifshitz, {\it Statistical Physics, 3rd Edition, Part 1}, Pergamon, Oxford, 1980, p. 160.
\bibitem{Chen} G. Chen, {\it Nanoscale energy Transport and Conversion}, Oxford University Press, 2005, Ch. 7.
\bibitem{Zhang} Z. M. Zhang, {\it Nano/Microscale Heat Transfer}, McGraw-Hill, New York, 2007, Ch. 5.
\bibitem{Ziman} J. M. Ziman, {\it Electrons and Phonons}, Oxford University Press, 1960.


\end{thebibliography}
\end{document}